\documentstyle[12pt,amsmath,amssymb,epsf,epic,cite]{article}

\numberwithin{equation}{section}

\newcommand{\e}{{\rm e}}

\renewcommand{\d}{{\rm d}}

\newcommand{\ed}{E_{\rm D}}
\newcommand{\ea}{E_{\rm \! A}}
\newcommand{\gd}{g_{\rm D}}
\newcommand{\ga}{g_{\rm \! A}}
\newcommand{\na}{N_{\rm \!A}}
\newcommand{\nd}{N_{\rm D}}
\newcommand{\sa}{\sigma_{\rm \!A}}
\newcommand{\sd}{\sigma_{\rm D}}

\newcommand{\Xd}{X_{\rm D}}
\newcommand{\Xa}{X_{\rm \! A}}
\newcommand{\xd}{x_{\rm D}}
\newcommand{\xa}{x_{\rm \! A}}
\newcommand{\pd}{p_{\rm D}}
\newcommand{\pa}{p_{\rm \! A}}

\newcommand{\diag}{{\rm diag}}

\newcommand{\av}[1]{\left\langle{#1}\right\rangle}

\newcounter{resultcounter}[section]

\newtheorem{thm}[resultcounter]{Theorem}

 \def\cE{{\cal E}} 
\def\cG{{\cal G}} \def\cH{{\cal H}}

\newcommand{\r}{{\rm R}}
\newcommand{\s}{{\rm S}}

\newcommand{\rx}{{\mathbb R}}
\renewcommand{\i}{{\rm i}}

\newcommand{\fer}[1]{(\ref{#1})}
\newcommand{\scalprod}[2]{\left\langle {#1}, {#2}\right\rangle}
\newcommand{\bbbone}{\mathchoice {\rm 1\mskip-4mu l} {\rm 1\mskip-4mu l}
{\rm 1\mskip-4.5mu l} {\rm 1\mskip-5mu l}}

\begin{document}

\title{\vspace*{-2cm} Electron Transfer Reactions:\\
Generalized Spin-Boson Approach}

\author{M. Merkli\footnote{Email: merkli@mun.ca, http://www.math.mun.ca/$\sim$merkli/ } \\
{\small Dept. of Mathematics and Statistics}\\ {\small  Memorial
University of Newfoundland} \\ {\small St. John's, NL, Canada A1C
5S7} \and  \medskip G.P.
Berman\footnote{Email: gpb@lanl.gov}\\
{\small Theoretical Division, MS B213}\\
{\small  Los Alamos National
Laboratory}\\
{\small  Los Alamos, NM 87545, USA}
\and \ R. Sayre\footnote{Email:  rsayre@newmexicoconsortium.org}\\
{\small Los Alamos National Laboratory and}\\
{\small New Mexico Consortium, 202B Research Center}\\
{\small Los Alamos, NM 87544, USA}\\
}
\date{\today}
\maketitle

\hfill LA-UR-12-24400

\begin{abstract}
We introduce a mathematically rigorous analysis of a generalized spin-boson system for the treatment of a donor-acceptor (reactant-product) quantum system coupled to a thermal quantum noise. The donor/acceptor probability dynamics describes transport reactions in chemical processes in presence of a noisy environment -- such as the electron transfer in a photosynthetic reaction center.  Besides being rigorous, our analysis has the advantages over previous ones that (1) we include a general, {\em non energy-conserving system-environment interaction}, and that (2) we allow for the donor or acceptor to consist of {\em multiple energy levels} lying closely together. We establish explicit expressions for the rates and the efficiency (final donor-acceptor population difference) of the reaction. In particular, we show that the rate increases for a multi-level acceptor, but the efficiency does not.
\end{abstract}

\bigskip
\bigskip

\thispagestyle{empty}
\setcounter{page}{1}
\setcounter{section}{1}
\setcounter{section}{0}

\section{Introduction}

\subsection{Transfer reactions and spin-boson model}


An important problem in chemistry and biology is to find electron transfer rates and transfer efficiencies in chemical reactions. A prominent example is the electron transfer in proteins carrying out photosynthesis (\!\!\cite{CF,OSSK,RMKLA,CBMTB}). The simplest reactions are described by two states, a reactant (electron donor) and a product (electron acceptor). Before the reaction, the system is localized mainly in the reactant state, and after mainly in the product state. The passage from reactant to product is induced by two effects: a {\em direct tunneling} (hopping) and an {\it indirect transition}. The former originates from electron tunneling between reactant and product moieties, while the latter is due to the presence of {\it thermal noise} created by the many protein atoms and molecules in which the electron donor are acceptor are embedded.

Denoting the reactant and product states by $|{\rm R}\rangle$ and $|{\rm P}\rangle$, respectively, a ``Marcus model'' Hamiltonian for the electron exchange has been used in \cite{Marcus1,XuSch},
$$
H_{\rm Marcus} = |{\rm R}\rangle E_{\rm R} \langle{\rm R}| +  |{\rm P}\rangle E_{\rm P} \langle{\rm P}| + |{\rm R}\rangle V \langle{\rm P}| + |{\rm P}\rangle V \langle{\rm R}|,
$$
where $E_{\rm R}$ and $E_{\rm P}$ are the reactant and product energies, and $V$ is the direct tunneling constant. Both $E_{\rm R}$ and $E_{\rm P}$ represent the collective energies of many particles (atoms and the molecules), corresponding to the reactant and product states of the protein environment. In the Marcus theory, the energy curves are taken to be harmonic in the collective position coordinate $q$,
$$
E_{\rm R} = \textstyle\frac12 fq^2,\qquad E_{\rm P}=\textstyle\frac12 f(q-q_{\rm P})^2-\epsilon_0.
$$
Here, $f$ denotes the common force constant of reactant and product, $q_{\rm P}$ is the equilibrium position of the product collective position (the reactant one being centered at the origin), and $\epsilon_0$ is the product-reactant energy difference. When describing the collective degrees of freedom of the environment quantum mechanically, the reactant and product energies become the operators with the Hamiltonians of a collection of harmonic oscillators \cite{XuSch}
$$
H_{\rm R} = \sum_\alpha \left(\frac{p_\alpha^2}{2m_\alpha} +\textstyle \frac12 m_\alpha\omega_\alpha^2q_\alpha^2\right), \quad
H_{\rm P} = \sum_\alpha \left(\frac{p_\alpha^2}{2m_\alpha} +\textstyle \frac12 m_\alpha\omega_\alpha^2(q_\alpha-q_{0,\alpha})^2 -\epsilon_{0,\alpha} \right).
$$
The number of oscillators corresponds to the number of atoms in the proteins and is very large (of the order of $10^4$ and more for the photosynthetic reaction center). Using these expression in the Hamiltonian $H_{\rm Marcus}$ above, we may write
$$
H_{\rm Marcus}=\left[
\begin{array}{cc}
H_{\rm R} & V\\
V & H_{\rm P}
\end{array}
\right],
$$
which, under proper identification of parameters (see \cite{XuSch}), has the form of the {\it spin-boson Hamiltonian} (plus a constant term which we drop)
\begin{equation}
H_{\rm SB} =  V\sigma_x + \textstyle\frac12\epsilon\sigma_z +H_{\rm R} +\lambda\sigma_z\varphi(h).
\label{sbh}
\end{equation}
Here, $\sigma_x$ and $\sigma_z$ are Pauli matrices, and
$$
\varphi(h) = \frac{1}{\sqrt{2}} \sum_\alpha h_\alpha (a^\dagger_\alpha+a_\alpha)
$$
is the bosonic ``field operator'', expressed in terms of the creation and annihilation operators satisfying $a_\beta a^\dagger_\alpha -a^\dagger_\alpha a_\beta =0$ if $\alpha\neq\beta$ and  $a_\alpha a^\dagger_\alpha -a^\dagger_\alpha a_\alpha=1$.

One achievement of Xu and Schulten's work \cite{XuSch} is the identification of the Marcus theory model with a spin-boson system. This identification allows to take over results obtained previously for the spin-boson system. In particular, Xu and Schulten use Leggett et al.'s \cite{Leggett} expression for the transfer rate in the spin-boson system, and show that at high temperatures, it coincides with the transfer rate predicted by the Marcus theory. In this setting, it is assumed that the direct coupling $V$ is very small. The term $V\sigma_x$ is then viewed as a perturbation of the other terms in the spin-boson Hamiltonian \fer{sbh}.

\medskip

We recognize that the works cited above are of great importance in the field. Our aim for the present paper is to try to improve some points.

\begin{itemize}
\item[1.] {\em Shortcoming.\ } The derivation of the transfer rate for small $V$ given in \cite{Leggett}, and then used by \cite{XuSch}, has two weak points. (1) The dominant term of the transfer rate is determined only heuristically (c.f. (3.31) of \cite{Leggett}, errors in the perturbation expansion, i.e., terms of order $V^2$ and higher, cannot be estimated). (2) Additional approximations (of Born-Markov type) are made in the derivation. Their validity has not been (cannot be) verified rigorously (see before (3.32) of \cite{Leggett}).

{\em Remedy.\ } We use the rigorous `dynamical resonance method' \cite{MBS,MSB} to find the dynamics of the reduced spin density matrix at all times $t\geq 0$, for arbitrary tunneling matrix elements $V$ and arbitrary energy separations $\epsilon$. By rigorous, we mean that our results are derived by a mathematical perturbation theory in the coupling $\lambda$ between spin and bath, in which the remainder terms are controlled and are small uniformly for all times $t\geq 0$.

\item[2.] {\em Shortcoming.\ } In the model \fer{sbh} used by \cite{XuSch}, the interaction term $\lambda\sigma_z\varphi(h)$ commutes with the main term $\frac12\epsilon\sigma_z$ (as $V$ is assumed to be small). This means that in absence of direct hopping ($V=0$), there is {\em no} electron transport at all. This is so since the populations are constant in time (diagonals of the density matrix in the energy representation). However, in reality, one would still expect electron transport due to the thermal noise, even if there is no direct hopping \cite{nalbach2012}.

{\em Remedy.\ } We modify the Hamiltonian \fer{sbh} by adding to the right side the term $\lambda a\sigma_x\varphi(h)$ which induces population dynamics (electron transport) even if $V=0$. Our analysis remains rigorous in presence of this term.

\item[3.] {\em Shortcoming.\ } In the above model, both the electron donor and acceptor are assumed to have {\em single levels}. However, due to the complexity of the biological system at hand, it is reasonable to consider that either (or both) of donor and acceptor consist of a number of levels $\nd$, $\na$, centered around an average energy $\ed$, $\ea$.

{\em Remedy.\ } We generalize the single-level system to the multi-level situation. We show that the multi-level model reduces to a single-level model with {\em rescaled Hamiltonian matrix elements}.
\end{itemize}

While our approach allows for the above-mentioned improvements, we can only treat small values of the coupling between the donor-acceptor system and the thermal reservoir. Throughout this work, we assume that
\begin{equation}
|\lambda| <\!\!< \ed-\ea.
\label{smlam}
\end{equation}
In this inequality, we consider $\lambda$ to be renormalized as to have the dimensionality of energy (i.e., \fer{smlam} means that $C |\lambda|< \ed-\ea$, where $C$ is a small constant depending on the vairous parameters of the model, and which is such that the left side has the dimensionality of energy).

The derivation of the electron transfer rate for small $V$ in \cite{Leggett} is based on the fact that the spin-boson Hamiltonian, for $V=0$, can be explicitly diagonalized (as in their model, the bath interaction commutes with the system Hamiltonian). As a result, the (heuristic) perturbation theory in $V$ of \cite{Leggett} yields an expression for the electron transfer rate which contains {\it all orders in} the spin-reservoir coupling $\lambda$. This approach cannot be carried out as soon as the interaction between spins and bosons is not proportional to $\sigma_z$, as the resulting Hamiltonian is not explicitly diagonalizable, even for $V=0$. Instead of containing all orders in $\lambda$ and second order in $V$ only, in our approach we obtain transfer rates (and the dynamics in general) to second order in $\lambda$ but to all orders in $V$, even in presence of a (de-)coherence altering interaction. We show in Section \ref{recleggsect} that the order $\lambda^2$ term of Leggett et al.'s transfer rate coincides with that obtained by our method.

\subsection{Transfer rates, separation}

Assume that initially, the donor is populated with probability one. In the course of time, the acceptor gains some population probability and for large times, the whole system converges to an asymptotic state. We call the {\em transfer rate} the speed at which the acceptor is populated, and the {\em transfer separation} the difference of acceptor minus donor population probability in the asymptotic state (after a long time). The results presented here are immediate consequences of a much stronger result, Theorem \ref{theorem1'} of Section \ref{sectmainres}, which gives the dynamics of the entire donor-acceptor density matrix, for all times.

\subsubsection{Single-level donor and acceptor}

The donor-acceptor-environment Hamiltonian is given by
\begin{equation}
H = \left[
\begin{array}{cc}
\ed & V\\
V & \ea
\end{array}
\right] +H_{\rm R} +\lambda \left[
\begin{array}{cc}
\gd & a\\
a & \ga
\end{array}
\right]\otimes\varphi(h).
\label{n0}
\end{equation}
Here, the first matrix is called the system Hamiltonian, it is the isolated donor-acceptor Hamiltonian, determined by the donor and acceptor energies, $\ed$ and $\ea$ (with $\ed>\ea$), and the tunnelling matrix element, $V\in\mathbb R$.  $H_{\rm R}$ is the Hamiltonian of the uncoupled reservoir, a field of harmonic oscillators
$$
H_{\rm R} = \sum_\alpha \omega_\alpha a^\dagger_\alpha a_\alpha,
$$
where we put $\hbar=1$, and $a_\alpha$, $a^\dagger_\alpha$ are bosonic annihilation and creation operators. We take the oscillators to be in thermal equilibrium at temperature $1/\beta>0$; they form a heat bath. An infinite-volume, or continuous-mode limit is taken in which the parameter $\alpha$ becomes the continuous boson momentum $k\in\rx^3$, see Section \ref{sectbathcorr} for more detail. The coupling between the donor-acceptor system and the bosonic reservoir is described by the third part on the right side of \fer{n0}. $\lambda\in\mathbb R$ is a coupling constant, $\gd,\ga\in\mathbb R$ and $a\in\mathbb R$ are interaction parameters responsible for energy conserving and energy exchange interactions. For $a=0$ (and when $V=0$) we have the purely energy-conserving interaction (the situation considered in \cite{XuSch}).

\medskip

Key dynamical properties depend on a few system (donor-acceptor) and reservoir (heat bath) quantities which we introduce now. The difference between the two eigenvalues of the system Hamiltonian is
\begin{equation}
\Omega=\sqrt{(\ed-\ea)^2+4V^2}\geq 0.
\label{dd1}
\end{equation}
Define the parameter $\alpha$, which can be positive, negative or zero, by
\begin{equation}
\alpha =\frac{V}{\ed-\ea}.
\label{dd2}
\end{equation}
The reservoir spectral density $J(\omega)$ is given, for $\omega\geq 0$, by
$$
J(\omega) = \sqrt{\pi/2} \tanh(\beta\omega/2) \left[\widehat{C}(\omega) + \widehat{C}(-\omega)\right],
$$
where $\widehat{C}(\omega)$ is the Fourier transform of
\begin{equation}
C(t)=\av{\e^{\i tH_{\rm R}}\varphi(h) \e^{-\i tH_{\rm R}}\varphi(h)}_\beta,
\label{corfun}
\end{equation}
the correlation function in the reservoir state at inverse temperature $\beta$. See Section \ref{sectbathcorr} for details.

\medskip

{\bf Transfer rates.} The decay of the donor and acceptor populations, $\pd$ and $\pa$, is a complicated function of time in general (see Theorem \ref{theorem1'}). However, in the regimes where either $\ed-\ea$ or $V$ is small relative to the other one, the population decay (growth) is exponential in time, and we can identify a transfer rate.
\begin{itemize}
\item[1.] For $|V|<\!\!<\ed-\ea$, we have
\begin{equation}
\pa(t) = \frac{1-\e^{\i t\varepsilon_0}}{1+\e^{-\beta\Omega}} +O(\lambda^2 +V),
\end{equation}
where $\varepsilon_0$ is a {\em complex (resonance) energy} (see \fer{dd3}).
The remainder is uniform (homogeneous) in $t\geq 0$. The acceptor is populated monotonically exponentially, at the rate $\gamma_{\rm relax}={\rm Im}\varepsilon_0$, which has the form
\begin{eqnarray}
\gamma_{\rm relax} &=& 2\lambda^2 \left[ \left(a-V \frac{\gd-\ga}{\ed-\ea}\right)^2 -4a^2V^2\right]\coth(\beta\Omega/2) J(\Omega)\nonumber\\
&&+O(\lambda^2V^3+\lambda^4).
\label{++5}
\end{eqnarray}
This result is obtained by taking $\alpha\rightarrow 0$ in Theorem \ref{theorem1'}.

\item[2.] For $\ed-\ea<\!\!<|V|$, we have
\begin{equation}
\pa(t) = \frac12 (1- {\rm Re\ } \e^{\i t\varepsilon_\Omega})+O(\lambda^2 +\ed-\ea),
\end{equation}
where $\varepsilon_\Omega$ is a {\em complex (resonance) energy} (see \fer{dd4}). The remainder is uniform (homogeneous) in $t\geq 0$. The acceptor is populated exponentially, but modulated by $\cos(t{\rm Re\ }\varepsilon_\Omega)$, with rate $\gamma_{\rm relax}'={\rm Im}\varepsilon_\Omega$, which has the form
\begin{eqnarray}
\lefteqn{
\gamma_{\rm relax}'}\nonumber\\
 &=& \lambda^2 \sqrt{\frac{\pi}{2}}\Big[ (\ga+\gd-2a)^2 -(\ed-\ea)(2+\frac{\ga-\gd}{V})(\ga+\gd-2a)\nonumber\\
&&+(\ed-\ea)^2\big[\frac{(\ga-\gd+a)(\ga+\gd-2a)}{2V^2} +(1+\frac{\ga-\gd}{2V})^2\big]\Big]\widehat C(0)\nonumber\\
&& +\frac{\lambda^2}{4}\left[\left( \ga-\gd-\frac{a(\ed-\ea)}{2V}\right)^2-\frac{(\ed-\ea)^2(\ga-\gd)^2}{4V^2}\right]\nonumber\\
&&\ \ \times\coth(\beta\Omega/2) J(\Omega)  +O\big(\lambda^2(\ed-\ea)^3+\lambda^4\big).
\end{eqnarray}
This result is obtained by taking $\alpha\rightarrow \infty$ in  Theorem \ref{theorem1'}.
\end{itemize}

{\bf Remark.\ } The relaxation rates \fer{+5}, \fer{gammaprime} contain the product 
$$
\coth(\beta\Omega/2)J(\Omega) = \sqrt{\pi/2} \left[\widehat{C}(\Omega) +\widehat{C}(-\Omega)\right],
$$
which depends on the temperature $1/\beta$ only via the bath correlation function $\widehat{C}$ (see after \fer{dd2}). We present the relaxation rates above involving the spectral density as this is customary in the literature.

\medskip
{\bf Transfer separation.} We define the separation by
$$
S=\pa(\infty)-\pd(\infty).
$$
It measures how much the populations are separated after relaxation and satisfies $-1\leq S\leq 1$. The extreme cases $S=\pm 1$ correspond to complete localization of the final state in level one or two. $S=0$ means complete delocalization (both levels equally probable). We obtain from Theorem \ref{theorem1'}
\begin{equation}
S = -1+\frac{2}{\sqrt{1+4\alpha^2}}\left\{ \frac{1}{1+\e^{-\beta\Omega}}+\frac{2\alpha^2}{1+\sqrt{1+4\alpha^2}}\right\}.
\label{sep}
\end{equation}
The separation does not depend on the initial state of the system.
For $\ed-\ea <\!\!<V$ (i.e., $\alpha$ large) we have $S\approx 0$, independently of the temperature. For $V<\!\!<\ed-\ea$ ($\alpha$ small) we have $S\approx -1+\frac{2}{1+\e^{-\beta\Omega}}=\tanh(\beta\Omega/2)$, which  becomes $S\approx 0$ at high temperatures ($\beta\rightarrow 0$), and $S\approx 1$ at low temperatures ($\beta\rightarrow\infty$). We conclude:

\medskip
{\em For large hopping constant ($\ed-\ea<\!\!<V$) the donor and acceptor are populated equally in the long run ($S\sim 0)$. The same happens for small hopping constant ($V<\!\!<\ed-\ea$) at high temperature. However, for small hopping constant ($V<\!\!<\ed-\ea$) and low temperature, the acceptor is fully populated in the long run (and the donor has probability zero).}

\subsubsection{Multi-level acceptor model}
\label{mlam}

{}For an $\na$-fold degenerate acceptor, the total Hamiltonian is
\begin{equation}
H = \left[
\begin{array}{c|ccc}
\ed & V     &  \cdots       & V \\
\hline
V    & \ea  &       &  \\
\vdots    &  &   \ddots    &    \\
V    &        &    & \ea
\end{array}
\right]
+ H_\r
 +\lambda
\left[
\begin{array}{c|ccc}
\gd & a     &  \cdots       & a \\
\hline
a    & \ga  &       &  \\
\vdots    &  &  \ddots    &  \\
a    &      &       & \ga
\end{array}
\right]\otimes\varphi(h).
\label{n1'}
\end{equation}
We introduce the donor state, $\varphi_{\rm D}$, and the {\em collective acceptor state}, $\sa$,
\begin{equation}
\varphi_{\rm D} =
\left[
\begin{array}{c}
1\\
0\\
\vdots\\
0
\end{array}
\right],\qquad
\sa=\frac{1}{\sqrt{\na}}\left[
\begin{array}{c}
0\\
1\\
\vdots\\
1
\end{array}
\right].
\label{n3'}
\end{equation}
In the basis $\{\varphi_{\rm D},\sa\}$, \fer{n1'} takes the form
\begin{equation}
H = \left[
\begin{array}{cc}
\ed & V\sqrt{\na}\\
V\sqrt{\na} & \ea
\end{array}
\right] +H_\r +\lambda
\left[
\begin{array}{cc}
\gd & a\sqrt{\na}\\
a\sqrt{\na} & \ga
\end{array}
\right]\otimes\varphi(h).
\label{n6'}
\end{equation}
This Hamiltonian is of the form \fer{n0}, {\em but in a different basis}, and {\em with rescaled off-diagonal matrix elements}. We can thus take over results obtained for the two-level model (see Section \ref{sectmainres} for details). As explained below, the final populations depend on the initial density matrix in the multi-level case. In this section, we assume that initially, the donor is fully populated and there is no donor-acceptor entanglement initially. In other words, the initial donor-acceptor density matrix is $|\varphi_{\rm D}\rangle\langle\varphi_{\rm D}|$.

\smallskip
{\bf Remark.\ } We consider here $V$ independent of $\na$, so the direct donor-accep\-tor interaction is of the size $V\na$. One may compensate this growth by scaling $V$ with a negative power of $\na$. We do not pursue this question in the present manuscript.

\bigskip
{\bf Transfer rates, multi-level acceptor model.} As in the case of the two-level system, the decay of populations is exponential in two limiting cases.
\begin{itemize}
\item[1.] For $\sqrt{\na}|V|<\!\!<\ed-\ea$, we have
\begin{equation}
\pa(t) =\frac{1}{\na} \frac{1-\e^{\i t\varepsilon_0}}{1+\e^{-\beta\Omega}}+O(\lambda^2 +V^2),
\end{equation}
where $\varepsilon_0$ is a complex resonance energy, depending on $\na$ (see Theorem \ref{thm3}). The acceptor is thus populated exponentially quickly and monotonically, at the rate
\begin{eqnarray}
\gamma_{\rm relax} &=& 2\na\lambda^2 \left[ \left(a-V \frac{\gd-\ga}{\ed-\ea}\right)^2 -4\na a^2V^2\right]\coth(\beta\Omega/2) J(\Omega)\nonumber\\
&& +O(\lambda^2V^3+\lambda^4).
\label{+5}
\end{eqnarray}
A proof of this is obtained by taking $\alpha$ small in Theorem \ref{thm2}.

\item[2.] For $\ed-\ea<\!\!<\sqrt{\na} |V|$, we have
\begin{equation}
\pa(t) = \frac{1}{2\na}(1-{\rm Re\ } \e^{\i t\varepsilon_\Omega})
+O(\lambda^2 +\ed-\ea),
\end{equation}
where $\varepsilon_\Omega$ is a complex resonance energy, depending on $\na$ (see Theorem \ref{thm3}).
The acceptor is populated exponentially, modulated by $\cos(t{\rm Re\ }\varepsilon_\Omega)$, with rate
\begin{eqnarray}
\gamma_{\rm relax}' &=& \lambda^2 \sqrt{\frac{\pi}{2}}\Big[ (\ga+\gd-2a\sqrt{\na})^2\nonumber\\
&& -(\ed-\ea)(2+\frac{\ga-\gd}{V\sqrt{\na}})(\ga+\gd-2a\sqrt{\na})\nonumber\\
&&+(\ed-\ea)^2\big[\frac{(\ga-\gd+a\sqrt{\na})(\ga+\gd-2a\sqrt{\na})}{2V^2\na}\nonumber\\
&& +(1+\frac{\ga-\gd}{2V\sqrt{\na}})^2\big]\Big]\widehat C(0)\nonumber\\
&& +\frac{\lambda^2}{4}\left[\left( \ga-\gd-\frac{a(\ed-\ea)}{2V}\right)^2-\frac{(\ed-\ea)^2(\ga-\gd)^2}{4V^2 \na}\right]\nonumber\\
&&\ \ \times\coth(\beta\Omega/2) J(\Omega)  +O\big(\lambda^2(\ed-\ea)^3+\lambda^4\big).
\label{gammaprime}
\end{eqnarray}
A proof of this is obtained by taking $\alpha$ small in Theorem \ref{thm2}. {}For large $\na$, we obtain
\begin{equation}
\gamma_{\rm relax}'\approx 2\sqrt{2\pi}\lambda^2a^2\na \widehat{C}(0) +\frac{\lambda^2}{4}\left(\ga-\gd-\frac{a(\ed-\ea)}{2V}\right)^2J(2V\sqrt{\na}).
\label{+++6}
\end{equation}
(Note that $\Omega\approx 2V\sqrt{\na}$ in this case.)
\end{itemize}

\bigskip

{\bf Transfer separation for the multi-level system.} Since all acceptor levels are populated equally at all times, we define the transfer separation for the multi-level acceptor system  by
\begin{equation}
S=\na \pa(\infty)-\pd(\infty).
\label{se2}
\end{equation}
Then $S=2\av{|\sa\rangle\langle\sa|}_\infty-1$, which is given again by \fer{sep}, but with
\begin{equation}
\alpha = \sqrt{\na} \frac{V}{\ed-\ea}.
\label{se1}
\end{equation}
For large $\na$, we have $S=0$, which means that all donor levels share probability $1/2$ and all acceptor levels share probability $1/2$ as well.

\bigskip

{\bf Discussion: effects of multiple levels.} {\bf (1)} {\em Transfer rate and degeneracy.\ } The effect of the $\na$-fold acceptor degereracy is to multiply the hopping coefficient $V$ and the decoherence coefficient $a$ by $\sqrt{\na}$, while it does not affect energy conserving parameters (see \fer{n6'}). Accordingly, the relaxation rates are (roughly) acquiring a factor $\na$ (as they are proportional to the square of the coefficients $V$ and $a$), see \fer{+5} and \fer{+++6}. Therefore, {\em $\na$-fold acceptor degeneracy speeds up the transfer process, the rate being proportional to $\na$}.

\smallskip

{\bf(2)} {\em Asymptotic population and degeneracy.\ } Due to the scaling, the Hamiltonian \fer{n6'} becomes, for large $\na$,
$$
 V\sqrt{\na}
\left[
\begin{array}{cc}
0 & 1\\
1 & 0
\end{array}
\right] +H_\r +\lambda a\sqrt{\na}
\left[
\begin{array}{cc}
0 & 1\\
1 & 0
\end{array}
\right]\otimes\varphi(h).
$$
Therefore, the off-diagonal density matrix elements of the donor-acceptor system (in the basis $\{\varphi_{\rm D},\sa\}$) is time-independent. Moreover, the system approaches the Gibbs equilibrium state in the long run, and the latter is, up to $O(\lambda^2)$-terms, equal to
$$
\frac{\e^{-\beta V\sqrt{\na}} |1\rangle_x\langle 1| +  \e^{-\beta V\sqrt{\na}} |-1\rangle_x\langle -1| }{2\e^{-\beta V\sqrt{\na}}} = \frac12 \bbbone.
$$
This is why, in presence of many acceptor levels, the final populations both in the donor, and in all acceptors together, equal one half each. Hence $\pa=1/(2\na)$.

\smallskip

{\bf (3)} {\em Separation and degeneracy.\ } In line with equal total distribution of donor and acceptor levels (previous point), the separation must vanish for large $\na$.

\smallskip

{\bf (4)} {\em Degenerate donor.\ } If the donor is $\nd$-fold degenerate and the acceptor is simple, then all the above formulas for the transfer rates and separations are the same, upon replacing $\na$ by $\nd$.

\smallskip

{\bf (5)} {\em Both acceptor and donor degenerate.\ } Some of our results hold if both the donor and the acceptor are degenerate. However, so far, we have not been able to find the dynamics of both the donor and acceptor in this setting, see the explanations in Section \ref{sectmainres}. However, a consideration as in point (2) above gives the following asymptotic result: if the donor and acceptor have degeneracies $\nd$ and $\na$, respectively, then the transfer rates scale as $\sqrt{\nd\na}$, and the separation becomes zero for large $\nd$ and $\na$. Asymptotically, each donor level has probability $1/(2\nd)$ and each acceptor level has probability $1/(2\na)$.

\smallskip

{\bf (6)} {\em Quasi-degenerate levels.\ } Our approach is also applicable if the levels are not exactly degenerate, but, say, spread around average values $\ed$ and $\ea$. More precisely, if, similar and in addition to \fer{smlam}, the spread $\Delta E$ of the levels satisfies $\Delta E<\!\!<\ed-\ea$, then the formulas below for transfer rates and separation give the correct lowest order terms in  $\Delta E$. In principle, one can calculate corrections of order $\Delta E$, but this is rather complicated.

\smallskip

{\bf (7)} {\em Dependence on initial condition.\ }  Consider a doubly-degenerate acceptor, system \fer{n1'} with 3$\times$3 matrices. This system has two invariant states (the kernel of $H$ has dimension two). One is immediately seen to be
$$
\tau=
\frac{1}{\sqrt{2}}
\left[
\begin{array}{c}
0\\
1\\
-1
\end{array}
\right]\otimes \Omega_{\rm R},
$$
where $\Omega_{\rm R}$ is the equilibrium state of the reservoir (satisfying $H_{\rm R}\Omega_{\rm R}=0$). The other stationary state is given by the 2D-reduced {\it Gibbs} equilibrium state of the two-dimensio\-nal system interacting with the reservoir, expressed in the basis $\{ \varphi_{\rm D},\sa\}$, \fer{n6'}. After tracing out the reservoir degrees of freedom, and to lowest order in the system-reservoir interaction $\lambda$, this 2D-Gibbs state is $\Omega_\s\propto\exp(-\beta H_\s)$, where $H_\s$ is the first matrix on the right side of \fer{n1'}. For instance, if $V=0$, then
\begin{equation}
\Omega_{\s} = \frac{\e^{-\beta \ed}|\varphi_{\rm D}\rangle\langle\varphi_{\rm D}|+\e^{-\beta\ea}|\sa\rangle\langle\sa|}{\e^{-\beta \ed}+\e^{-\beta \ea}}.
\label{ndgibbs}
\end{equation}
Note that in the original basis $\{\varphi_{\rm D}, \varphi_1,\varphi_2\}$, in which the matrices in \fer{n1'} are presented, the final state \fer{ndgibbs} {\it is not even diagonal}.

A general initial condition will converge, for large times, to a superposition of the two invariant states. If the initial condition belongs to one of the invariant subspaces ${\rm span}\{\varphi_{\rm D},\sa\}$ or ${\mathbb C}\tau$, then its final state will be the 2D-reduced Gibbs state or $\tau$. In the above results, we assume that the initial condition is the pure state entirely concentrated on the donor. It belongs to the subspace spanned by $\{\varphi_{\rm D},\sa\}$ and hence the final state is the associated Gibbs state. A similar dependence of the asymptotic state on the initial condition holds for any acceptor dimension. We note that a deviation of the Gibbs equilibrium as a final state of an open system has also been observed in \cite{WGZ}, in the setting of a spin coupled symmetrically to a bath of other spins in thermal equilibrium (`spin star system').

\smallskip

{\bf (8)} {\em Continuous acceptor, or sink, or Wigner-Weiskopf limit in presence of a heat bath?\ } In so-called ``sink'' or Wigner-Weiskopf models, a dissipative part of the system is modeled by considering a system of interest (the donor) coupled to $N$ energy levels (the acceptors), and the limit $N\rightarrow\infty$ is taken in order to obtain irreversible phenomena (decay). See, for example \cite{WW}, p. 36 and following. In taking the continuum limit, the typical spacing between individual levels, $\Delta E$, is decreased more and more and at the end the levels are characterized by a continuous {\em density of states}. The typical role of a sink is to depopulate the donor exponentially quickly (at a rate proportional to the density of states). One may ask if, in taking $\na\rightarrow\infty$ in our model, we obtain the same dynamical or asymptotic results as for a sink model. The answer is negative for the reason we explain below. However, we point out that the rigorous derivation of the Wigner-Weiskopf model has never been done (to our knowledge) in the presence of an additional heat bath.

The following transition occurs when taking the $\Delta E\rightarrow 0$ limit in presence of a thermal bath. Let $\lambda$ be the coupling strength of the donor-acceptor system with the thermal bath. If $\lambda <\!\!< \Delta E$, then the donor-acceptor energy levels ``are well defined'' (as without interaction with the heat bath). In this case, the asymptotic state is given by the Gibbs state of the donor-acceptor system ($\propto\exp(-\beta H_\s)$, modulo corrections small in $\lambda$). The final donor population is consequently (consider $V=0$ in \fer{n1'})
$$
\pd=\frac{\e^{-\beta\ed}}{\e^{-\beta\ed}+N \e^{-\beta\ea}},
$$
which decreases as $1/N$. In this regime the transfer has a good efficiency (almost total depopulation of the donor), as is the case in the Wigner-Weiskopf model. However, as we increase $N$, we decrease $\Delta E$ and we reach the regime $\lambda>\!\!>\Delta E$. This case is a perturbation of the totally degenerate situation ($\Delta E=0$) which we treat in the present paper. The 2D-reduction \fer{n6'} then takes place, and we obtain a final donor population (see after \fer{se1})
$$
\pd \approx 1/2,
$$
which means that the transfer is not very efficient (and cannot be made more so by increasing the number $N$ of acceptors).

To sum up: Our analysis holds for arbitrary $N$, but in the limit $N\rightarrow\infty$, it does not give depopulation of the donor. The depopulation however holds in sink-models without a thermal bath.

\section{Main results: details}
\label{sectmainres}

\subsubsection{Two-level donor-acceptor model}

The total Hamiltonian is given by \fer{n0}. We point out that $J(\omega)$ is {\it independent} of the temperature. It has the explicit representation ($\omega>0$)
$$
J(\omega) = \omega^2\int_{S^2}|h(\omega,\Sigma)|^2\d \Sigma,
$$
where the integral is over the two-dimensional sphere, and the function $h=h(k)$ is written in spherical coordinates for $k=(\omega,\Sigma)\in{\mathbb R}^3$. (In the last expression for $J$, the oscillators are indexed by $\alpha\rightarrow k$ in the continuum, or infinite volume limit for the momentum $k\in{\mathbb R}^3$.)

Let $\rho_t$ be the {\it reduced donor-acceptor density matrix}, when the degrees of freedom of the environment are traced over. We assume that initially, the entire system is in a state of the form
$$
\rho_{\rm in} = \rho_0\otimes\rho_{\rm R},
$$
where $ \rho_0$ is an arbitrary two-state initial density matrix, and $\rho_{\rm R}$ is the initial state of the reservoir, taken to be at equilibrium at inverse temperature $\beta$. Then we have the dynamical equation
$$
\rho_t = {\rm Tr}_{\rm R}\left[ \e^{-\i tH}\rho_{\rm in}\e^{\i tH}\right]
$$
for the reduced density matrix. Here, the trace is taken over the reservoir space.

The two eigenvalues of the donor-acceptor Hamiltonian are
$$
\frac 12\left[\ed+\ea \pm\sqrt{(\ed-\ea)^2+4V^2}\right],
$$
resulting in the eigenvalue difference \fer{dd1}. We define the `resonance energies' by
\begin{eqnarray}
\varepsilon_0 &=& 2\i \lambda^2 \frac{[a+(\ga-\gd)\alpha]^2}{1+4\alpha^2} \coth(\beta\Omega/2) J(\Omega)\label{dd3}\\
\varepsilon_\Omega &=& \varepsilon_0/2-\Omega -\lambda^2 X +\i\lambda^2\sqrt{\pi/2}\  Y^2 \widehat C(0) \label{dd4}
\end{eqnarray}
where
\begin{equation}
Y= (\ga-\gd)\frac{4\alpha^2}{\sqrt{1+4\alpha^2}\left(\sqrt{1+4\alpha^2}-1\right)} +2\gd -a\frac{4\alpha}{\sqrt{1+4\alpha^2}} +\ea-\ed.
\label{dd5}
\end{equation}
and
\begin{eqnarray}
X &=& (\ed-\ea)\left( 4\gd -a\frac{8\alpha}{\sqrt{1+4\alpha^2}}-Y \right){\rm Im}\int_0^\infty\av{\varphi(t)\varphi}_{\beta}\d t\nonumber\\
&& +2\sqrt{2/\pi} \frac{[a+(\ga-\gd)\alpha]^2}{1+4\alpha^2} \ {\rm Re} \int_0^\infty \sin(\Omega t) \av{\varphi(t)\varphi}_\beta \d t.
\label{dd6}
\end{eqnarray}
Here, $\av{\ }_\beta$ denotes the average in the thermal equilibrium  of the environment. We denote by $[\rho_t]_{ij}$ the matrix elements of the reduced density matrix of the two-level system in the basis $\{[1\ 0]^t, [0\ 1]^t\}$. The following is our main result. It describes the population dynamics of the two-level system, identifying a main part and a remainder, which is of order $O(\lambda^2)$, {\it homogeneously} in (independent of) time.

\begin{thm}
\label{theorem1'}
The dynamics of the product (acceptor) probability $[\rho_t]_{22}$, for $t\geq 0$, is given by
\begin{eqnarray}
[\rho_t]_{22} &=& [\rho_0]_{22}\frac{1}{1+4\alpha^2} \left\{ \e^{\i t\varepsilon_0}+4\alpha^2{\rm Re}\ \e^{\i t\varepsilon_\Omega}\right\}\nonumber\\
&&-{\rm Re}\left([\rho_0]_{12}\right)\frac{2\alpha}{1+4\alpha^2}\left\{ \e^{\i t\varepsilon_0} -{\rm Re}\ \e^{\i t\varepsilon_\Omega}\right\}\nonumber\\
&& +{\rm Im}\left([\rho_0]_{12}\right)\frac{2\alpha}{1+4\alpha^2} {\rm Im}\ \e^{\i t\varepsilon_\Omega}\nonumber\\
&&-\frac{[1+\e^{-\beta\Omega}]^{-1}}{\sqrt{1+4\alpha^2}}\left(\e^{\i t\varepsilon_0}-1\right) +\frac{\e^{\i t\varepsilon_0}+\sqrt{1+4\alpha^2}}{1+\sqrt{1+4\alpha^2}}\frac{2\alpha^2}{1+4\alpha^2}\nonumber\\
&& -\frac{2\alpha^2}{1+4\alpha^2}{\rm Re}\ \e^{\i t\varepsilon_\Omega} +O(\lambda^2).
\label{f3}
\end{eqnarray}
The remainder term $O(\lambda^2)$ is independent of $t\geq 0$ (and of $\ed,\ea,V,a$ varying in bounded sets).
\end{thm}

{\em Remark.\ } In the ``usual'' setup \cite{MSB,MBS,MEVol} for the derivation of the reduced density matrix \fer{f3}, we start with a diagonal system Hamiltonian. Then the diagonal density matrix elements evolve jointly, and only the `resonances bifurcating out of the zero eigenvalue' ($\varepsilon_0$ here) are present in their evolution. However, in the present setup, $H_\s$ is not diagonal, and as a result, the evolution of the diagonal involves the initial condition of the {\em off-diagonal} density matrix elements, and the evolution also depends on the resonances bifurcating out of the non-zero eigenvalues ($\varepsilon_{\Omega}$ here).

\subsubsection{Multi-level donor-acceptor model}
\label{symmsect}

We consider an $\nd$-fold donor and an $\na$-fold acceptor with energies $\ed$ and $\ea$, respectively. The energy levels may be distributed around these two fixed energies, provided their spread is small. The total Hamiltonian is
\begin{eqnarray}
H &=& \left[
\begin{array}{ccc|ccc}
\ed &        &     & V     &  \ldots       & V \\
    & \ddots &     & \vdots      &       &  \vdots   \\
    &        & \ed & V     &   \ldots      & V \\
\hline
V    &     \ldots     & V    & \ea  &       &  \\
\vdots    &        &  \vdots  &      &\ddots    \\
V    &   \ldots     & V    &      &       & \ea
\end{array}
\right]
+ H_\r \nonumber\\
&& +\lambda
\left[
\begin{array}{ccc|ccc}
\gd &        &     & a     &  \ldots       & a \\
    & \ddots &     & \vdots      &       &  \vdots   \\
    &        & \gd & a     &   \ldots      & a \\
\hline
a    &     \ldots     & a    & \ga  &       &  \\
\vdots    &        &  \vdots  &      &\ddots    \\
a    &   \ldots     & a    &      &       & \ga
\end{array}
\right]\otimes\varphi(h).
\label{n1}
\end{eqnarray}
The donor-acceptor space is partitioned into $\nd$ levels $\ed$ and $\na$ levels $\ea$. By introducing the vectors
\begin{equation}
\sd=\frac{1}{\sqrt{\nd}}\left[
\begin{array}{c}
1\\
\vdots\\
1\\
0\\
\vdots\\
0
\end{array}
\right],\qquad
\sa=\frac{1}{\sqrt{\na}}\left[
\begin{array}{c}
0\\
\vdots\\
0\\
1\\
\vdots\\
1
\end{array}
\right],
\label{n3}
\end{equation}
the Hamiltonian \fer{n1} can be written as
\begin{eqnarray}
H &=& \cE +V\sqrt{\nd\na}\big\{ |\sd\rangle\langle\sa| + |\sa\rangle\langle\sd|\big\} +H_\r\nonumber\\
&& +\lambda \Big\{\cG +a\sqrt{\nd\na}\big( |\sd\rangle\langle\sa| + |\sa\rangle\langle\sd|\big)\Big\}\otimes\varphi(h),
\label{n2}
\end{eqnarray}
where
\begin{eqnarray}
\cE &=& \diag(\ed,\ldots,\ed,\ea,\ldots,\ea),\\
\cG &=& \diag(\gd,\ldots,\gd,\ga,\ldots,\ga).
\label{n5}
\end{eqnarray}

\bigskip

{\bf Reduction to two-level system.\ } From \fer{n2}-\fer{n5} we see that $H$ leaves the subspace
$$
\overline\cH = {\rm span}\{\sd,\sa\}\otimes\cH_\r
$$
invariant. This means that if $\chi\in\overline{\cH}$ then $H\chi\in\overline{\cH}$. It is the same as saying that $H$ can be written as a block-diagonal matrix in the decomposition
$$
\cH = \overline{\cH}\oplus {\overline\cH}^\perp.
$$
Of course, each block is still infinite-dimensional due to the reservoir degrees of freedom, but the block in $\overline{\cH}$ involves only {\it two} donor-acceptor vectors, namely $\sd$ and $\sa$. This is why the the multi-level donor-acceptor model has a two-level formulation. In the basis $\{\sd,\sa\}$, $H$ in \fer{n1} takes the form
\begin{equation}
H = \left[
\begin{array}{cc}
\ed & V\sqrt{\nd\na}\\
V\sqrt{\nd\na} & \ea
\end{array}
\right] +H_\r +\lambda
\left[
\begin{array}{cc}
\gd & a\sqrt{\nd\na}\\
a\sqrt{\nd\na} & \ga
\end{array}
\right]\otimes\varphi(h).
\label{n6}
\end{equation}
This Hamiltonian is of the form \fer{n0} with rescaled off-diagonal coefficients $V\rightarrow V\sqrt{\nd\na}$ and $a\rightarrow a\sqrt{\nd\na}$.

\bigskip

{\bf Symmetry.\ }
Due to the symmetry of the Hamiltonian $H$, \fer{n2}, the donor-acceptor density matrix has a special structure. For $1\leq i,j\leq \nd+\na$, let $U_{ij}$ be the unitary operator wich exchanges labels $i$ and $j$. In other words, $U_{ij}\varphi_i=\varphi_j$, $U_{ij}\varphi_j=\varphi_i$ and $U_{i,j}\varphi_k=\varphi_k$ if $k\neq i,j$, where $\{\varphi_k\}_{k=1}^{\nd+\na}$ is the energy basis (in which \fer{n1} and \fer{n3} are expressed).

We consider initial density matrices $\rho_0$ which are symmetric with respect to permutation within the donor and within the acceptor subspaces,
\begin{equation}
U_{ij}\, \rho_0\, U_{ij}=\rho_0
\label{n7}
\end{equation}
if $1\leq i,j\leq \nd$ and if $\nd+1\leq i,j\leq \nd+\na$. An example of a symmetric initial state is
\begin{equation}
\rho_0 = \frac{1}{\nd}{\rm diag}(1,\ldots,1,0,\ldots,0),
\label{n8}
\end{equation}
in which each donor degree of freedom is populated equally likely.

\begin{thm}[Symmetry]
\label{thm2}
Suppose that the initial state is symmetric as in \fer{n7}. The reduced density matrix of the donor-acceptor system has the form
$$
\rho_t = \left[
\begin{array}{c|c}
\Xd(t) &  *\\
\hline
 *  & \Xa(t)
\end{array}
\right],
$$
where the $*$ represent some matrices, and where $\Xd$ and $\Xa$ are square matrices of size $\nd$ and $\na$, of the form
$$
\Xd(t) = \left[
\begin{array}{cccc}
\pd & \xd &\cdots & \xd \\
\xd & \pd &\ddots & \vdots \\
\vdots & \ddots &\ddots & \xd\\
\xd & \cdots &\xd & \pd
\end{array}
\right],\qquad
\Xa(t) = \left[
\begin{array}{cccc}
\pa & \xa &\cdots & \xa \\
\xa & \pa &\ddots & \vdots \\
\vdots & \ddots &\ddots & \xa\\
\xa & \cdots &\xa & \pa
\end{array}
\right].
$$
The off-diagonal matrix elements of $\Xd$ are all equal and real, and so are those of $\Xa$. All diagonals of $\Xd$ are equal, and so are those of $\Xa$, and they satisfy $\pa\na+\pd\nd=1$.
\end{thm}

\bigskip

{\it Proof of Theorem \ref{thm2}.\ } For $1\leq i,j,k\leq \nd$ we have
\begin{eqnarray}
[\rho_t]_{ij}&=& {\rm Tr}\left( U_{kj}U_{kj}\rho_0 \,\e^{\i tH} |\varphi_j\rangle\langle\varphi_i| \e^{-\i tH}\right)\nonumber\\
&=& {\rm Tr}\left( \rho_0 \,\e^{\i tH} U_{kj}|\varphi_j\rangle\langle\varphi_i| U_{kj}\e^{-\i tH}\right). \label{n4}
\end{eqnarray}
If $i=j$ then $U_{kj}|\varphi_j\rangle\langle\varphi_i| U_{kj}=|\varphi_k\rangle\langle\varphi_k|$ and \fer{n4} means that $[\rho_t]_{ii}=[\rho_t]_{kk}$. If $i\neq j$ then: $U_{kj}\varphi_i=\varphi_j$ if $k=i$ (so \fer{n4} gives $[\rho_t]_{ij}=[\rho_t]_{ji}$) and $U_{kj}\varphi_i=\varphi_i$ if $k\neq i$ (in which case \fer{n4} gives $[\rho_t]_{ij}= [\rho_t]_{ik}$). This shows that $A$ is of the form as given in the theorem. Repeating the same argument for indices $\nd+1\leq i,j,k\leq \nd+\na$ yields the form of $B$. The relation between $p$ and $q$ is obtained from ${\rm Tr} \,\rho_t=1$. \hfill $\blacksquare$

\bigskip

Consider the average $\av{|\sa\rangle\langle\sa|}_t$, where we recall that $\sa$ is given in \fer{n3}. On the one hand, we have
\begin{eqnarray}
\av{|\sa\rangle\langle\sa|}_t &=&{\rm Tr}\left(\rho_0\e^{\i tH} |\sa\rangle\langle\sa| \e^{-\i tH}\right)\nonumber\\
&=& \frac{1}{\na}\sum_{i,j=\nd+1}^{\na+\nd} {\rm Tr}\left(\rho_0\e^{\i tH} |\varphi_j\rangle\langle\varphi_i| \e^{-\i tH}\right)\nonumber\\
&=& \frac{1}{\na}\sum_{i,j=\nd+1}^{\na+\nd} [\rho_t]_{ij}\nonumber\\
&=& \frac{\pa\na+ \xa(\na^2-\na)}{\na} = \pa+\xa(\na-1),
\label{n9}
\end{eqnarray}
where $\pa$ and $\xa$ are the matrix elements of the acceptor block given in Theorem \ref{thm2}. On the other hand, by comparing \fer{n6} with \fer{n0}, $\av{|\sa\rangle\langle\sa|}_t$ equals $[\rho_t]_{22}$ in the formalism of the two-level model. Theorem \ref{theorem1'} thus yields the following result.
\begin{thm}
\label{thm3} Suppose that initially, the donor degrees of freedom are populated only, with equal probability $1/\nd$, as in \fer{n8}. Then we have
\begin{eqnarray}
\av{|\sa\rangle\langle\sa|}_t &=& \frac{[1+\e^{-\beta\Omega}]^{-1}}{\sqrt{1+4\alpha^2}}\left(1-\e^{\i t\varepsilon_0}\right) +\frac{\e^{\i t\varepsilon_0}+\sqrt{1+4\alpha^2}}{1+\sqrt{1+4\alpha^2}}\frac{2\alpha^2}{1+4\alpha^2}\nonumber\\
&& -\frac{2\alpha^2}{1+4\alpha^2}{\rm Re}\ \e^{\i t\varepsilon_\Omega} +O(\lambda^2).
\label{n10}
\end{eqnarray}
Here, $\alpha, \Omega,\varepsilon_0, \varepsilon_\Omega$ are given as in \fer{dd1}-\fer{dd6}, but with $a$ and $V$ replaced by $a\sqrt{\nd\na}$ and $V\sqrt{\nd\na}$, respectively.
\end{thm}

By proceeding in the same way, one finds
\begin{equation}
\av{|\sd\rangle\langle\sd|}_t = \pd +\xd(\nd-1),
\label{n11}
\end{equation}
where $\pd$, $\xd$ define the donor block $\Xd$ defined in Theorem \ref{thm2}, and $\av{|\sd\rangle\langle\sd|}_t$ is equal to $[\rho_t]_{11}=1-[\rho_t]_{22}$ in the two-site model (Theorem \ref{theorem1'}). This means that  $\av{|\sd\rangle\langle\sd|}_t +\av{|\sa\rangle\langle\sa|}_t=1$, hence \fer{n9} and \fer{n11} give
\begin{equation}
\pa +\xa(\na-1) +\pd+\xd(\nd-1)=1.
\label{n12}
\end{equation}
Together with the equation $\pd\nd+\pa\na=1$ (see Theorem \ref{thm2}), \fer{n9}, \fer{n11} and \fer{n12} are four equations for four unknowns $\pa,\pd,\xa,\xd$. However, only three of those equations are independent (as \fer{n12} is the sum of \fer{n9} and \fer{n11}), so the solution is indetermined. In case $\na=1$ or $\nd=1$, one variable is eliminated (for instance, if $\nd=1$ then $\xd$ is not present), and the system of equations can be solved. We consider this next.

\bigskip

{\bf Single-level donor to multi-level acceptor.} We look at $\nd=1$ and $\na\geq 1$ arbitrary. Then
$$
\pd=1-\av{|\sa\rangle\langle\sa|}_t, \quad \pa=\frac{1}{\na}\av{|\sa\rangle\langle\sa|}_t,\quad \xa=\frac{1}{\na}\av{|\sa\rangle\langle\sa|}_t.
$$
The first equation comes from \fer{n11} and $\av{|\sd\rangle\langle\sd|}_t +\av{|\sa\rangle\langle\sa|}_t=1$. The second equation then follows from $\pd\nd+\pa\na=1$. Finally, the third equation comes from \fer{n12}.

\bigskip

{\bf Multi-level donor to single-level acceptor.} Here $\nd\geq 1$ and $\na=1$. Then
$$
\pd=\frac{1-\av{|\sa\rangle\langle\sa|}_t}{\nd}, \quad \pa=\av{|\sa\rangle\langle\sa|}_t,\quad \xd=\frac{1-\av{|\sa\rangle\langle\sa|}_t}{\nd}.
$$

The statements about the transfer rates and separation in Secton \fer{mlam} follow directly from the above formulas.

\section{Proof of Theorem \ref{theorem1'}}

The dynamical resonance method \cite{MBS,MSB} gives the reduced dynamics of the spin system (donor-acceptor system). Its starting point is a {\em diagonal} system Hamiltonian. Then a rigorous perturbation theory is applied, tracing out the reservoir degrees of freedom and describing how the system energies become complex (`resonance energies') and lead to decay. Those complex energies are the eigenvalues of a (non-hermitian) effective energy operator, called a ``level shift operator''. The task is thus to diagonalize the $2\times 2$ system Hamiltonian, and then to calculate and diagonalize the level shift operators. We do not carry out all details as this would take up too much space.

The system Hamiltonian
$$
H_\s = \left[
\begin{array}{cc}
\ed & V\\
V & \ea
\end{array}
\right]
$$
is diagonalized by the unitary
\begin{equation}
U= \left[
\begin{array}{cc}
\frac{V}{\sqrt{V^2+\zeta_1^2}} & \frac{\zeta_1}{\sqrt{V^2+\zeta_1^2}}\\
\frac{V}{\sqrt{V^2+\zeta_2^2}} & \frac{\zeta_2}{\sqrt{V^2+\zeta_2^2}}
\end{array}
\right],
\label{f1}
\end{equation}
where 
$$
\zeta_{1,2} = \frac12\{ \ed+\ea\mp\sqrt{(\ed-\ea)^2+4V^2}\}.
$$
In the new basis, the Hamiltonian \fer{n0} has the form
$$
\widetilde H = UH U^{-1} = \widetilde H_\s +H_\r +\lambda\widetilde W\otimes\varphi(h),
$$
where
$$
 \widetilde H_\s := UH_\s U^{-1} ={\rm diag}(\zeta_1,\zeta_2)
$$
and $\widetilde W = U W U^{-1}$. As the system Hamiltonian is now diagonal, we can apply the dynamical resonance theory \cite{MSB,MBS} to find the dynamics of the reduced system density matrix. We outline the most important steps.

In the Gelfand-Naimark-Segal Hilbert space representation of the system, the density matrices on the spin-boson Hilbert space, ${\mathbb C}^2\otimes{\cal F}(L^2(\rx^3,\d x^3)$, are identified with {\em vectors} of a new Hilbert space ${\cal H}_{\rm GNS} = {\mathbb C}^2\otimes {\mathbb C}^2\otimes {\cal F}(L^2(\rx\times S^2,\d u\times\d\Sigma)$. This ``doubling'' of the space is described in detail in \cite{MBS,MSB}. Here, ${\cal F}(X)$ is the {\em Fock} space over the one-particle space $X$ \cite{BR}. For $X=L^2(\rx\times S^2,\d u\times\d\Sigma)$ it carries the creation operators and annihilation operators $a(u,\Sigma), a^\dagger(u,\Sigma)$ satisfying $[a(u,\Sigma),a^\dagger(u',\Sigma')]=\delta(u-u')\delta(\Sigma-\Sigma')$ (Kronecker deltas). The {\em Liouville operator} is defined by
$$
\widetilde K = \widetilde L_\s + L_\r +\lambda\widetilde I,
$$
where $\widetilde L_\s=\widetilde H_\s\otimes\bbbone_\s - \bbbone_\s\otimes\widetilde H_\s$, $L_\r=\d\Gamma(u)$ is the second quantization of the operator of multiplication by the argument $u\in\rx$ in $L^2(\rx\times S^2)$. The operator $\widetilde I$ represents the interaction between the spin and the bosons. The explicit form is $\widetilde I = \widetilde W\otimes\bbbone_\s\otimes\varphi_\beta(h) -J\Delta^{1/2}(\widetilde W\otimes\bbbone_\s\otimes\varphi_\beta(h))J\Delta^{1/2}$, where $J,\Delta$ are the modular conjugation and the modular operator associated to the vector $\Omega_\s\otimes\Omega_\r\in{\cal H}_{\rm GNS}$. Here, $\Omega_\s$ is the trace state of the spin and $\Omega_\r$ is the vacuum vector, representing the equilibrium state at temperature $1/\beta>0$ of the infinitely extended bose gas. The modular data is a concept of {\em Tomita-Takesaki theory} of von-Neumann algebras. We do not define these objects here, but refer to \cite{MBS,MSB,BR} for details. The spectrum of $\widetilde L_\s$ consists of energy differences of $\widetilde H_\s$. The spectrum of $L_\r$ has a simple eigenvalue zero with eigenvector $\Omega_\r$ and is otherwise continuous, covering the whole real axis. The four eigenvalues of $\widetilde K_0=\widetilde L_\s+L_\r$ are $\pm(\zeta_1-\zeta_2)$ (each simple) and zero (twice degenerate). These eigenvalues are embedded in the continuous spectrum. The achievement of the dynamical resonance theory is to describe the instability of the eigenvalues under the perturbation $\lambda \widetilde I$. More precisely, these eigenvalues become the {\em complex} eigenvalues of a {\em spectrally deformed} version of the operator $\widetilde K$. To each of the eigenvalue $e$ of $\widetilde K_0$ is associated a ``level shift operator'' $\Lambda_e$. The eigenvalues of $\Lambda_e$ are (up to fourth-order corrections in $\lambda$) the complex eigenvalues. For $e=\pm(\zeta_1-\zeta_2)$, $\Lambda_e$ is simply a number, namely the operator acting on the one-dimensional space ${\mathbb C}\varphi_1\otimes\varphi_2\otimes\Omega_\r$ and ${\mathbb C}\varphi_2\otimes\varphi_1\otimes\Omega_\r$, where {\em $\varphi_{1,2}$ is the canonical basis in which $\widetilde H_\s$ is diagonal}. The operator $\Lambda_0$ is two-dimensional (as zero is doubly degenerate). We now give the explicit form of these level shift operators. (Their definition and calculation is rather straightforward, albeit somewhat lengthy -- we refer to \cite{MBS,MSB} for a general formulas for the level shift operators). We have
\begin{equation}
\Lambda_0=
\i\lambda^2 \sqrt{2\pi} \left|\scalprod{\varphi_1}{\widetilde W\varphi_2}\right|^2 
\left[
\begin{array}{cc}
\widehat C(\Omega) & -\widehat C(\Omega)\\
-\widehat C(-\Omega) & \widehat C(-\Omega)
\end{array}
\right],
\label{lnot}
\end{equation} 
where $\Omega$ is given in \fer{dd1} and where $\widehat C$ is the Fourier transform of the correlation function \fer{corfun} (see also Section \ref{sectbathcorr}). This operator is written in the basis $\{\varphi_1\otimes\varphi_1, \varphi_2\otimes\varphi_2\}$. The eigenvalues of \fer{lnot} are $0$ and $\varepsilon_0$ (see \fer{dd3}). The eigenprojection onto the eigenvalue zero is
$$
Q_0^{(0)}= 
\frac{1}{\widehat C(\Omega) + \widehat C(-\Omega)}
\left|
\left[
\begin{array}{c}
1 \\
1
\end{array}
\right]\right\rangle
\left\langle
\left[
\begin{array}{c}
\widehat C(-\Omega) \\
\widehat C(\Omega)
\end{array}
\right]\right|,
$$
while that on the eigenvalue $\varepsilon_0$ is 
$$
Q_0^{(1)}= \frac{-1}{\widehat C(\Omega) + \widehat C(-\Omega)}
\left|
\left[
\begin{array}{c}
1 \\
-1
\end{array}
\right]\right\rangle
\left\langle
\left[
\begin{array}{c}
-\widehat C(\Omega) \\
\widehat C(-\Omega)
\end{array}
\right]\right|.
$$
We point out the formula $\widehat C(\Omega) + \widehat C(-\Omega)= \Omega^2\coth(\beta\Omega/2) $, see Section \ref{sectbathcorr}. 
Similarly, one finds for the level shift operator associated to $\zeta_2-\zeta_1= \Omega$
\begin{equation}
\Lambda_{\Omega} = \varepsilon_\Omega \ |\varphi_2\otimes\varphi_1\rangle\langle\varphi_2\otimes\varphi_1|,
\label{l12}
\end{equation}
where $\varepsilon_\Omega$ is given in \fer{dd4}. Similarly, $\Lambda_{-\Omega} = -\overline{\varepsilon_\Omega} \ |\varphi_1\otimes\varphi_2\rangle\langle\varphi_1\otimes\varphi_2|$. 

This information is sufficient to give the dynamics of the reduced system density matrix $\widetilde\rho_t$ in the basis in which $\widetilde H_\s$ is diagonal (see e.g. Theorem 2.1 in \cite{MBS} or Theorem 2.1 in \cite{MEVol}). For example,
\begin{equation}
[\widetilde\rho_t]_{11}:=\scalprod{\varphi_1}{\widetilde\rho_\s\varphi_1}= A_t(11;11)[\widetilde\rho_0]_{11} +A_t(11;22)[\widetilde\rho_0]_{22} +O(\lambda^2),
\label{f2}
\end{equation}
where 
$$
A_t(11;kk)=\scalprod{\varphi_k\otimes\varphi_k}{Q_0^{(0)}\varphi_1\otimes\varphi_1} +\e^{\i t\varepsilon_0^{(1)}} \scalprod{\varphi_k\otimes\varphi_k}{Q_0^{(1)}\varphi_1\otimes\varphi_1}.
$$
The remainder term in \fer{f2} is uniform in $t\geq 0$. 
To obtain the dynamics of the reduced system state in the original basis (in which $H_\s$ is not diagonal), we undo the base-change implemented by $U$, according to $\rho_t=U^{-1}\widetilde\rho_t U$, c.f. \fer{f1}. This yields (after some algebra) the relation \fer{f3} for $[\rho_t]_{22}=1-[\rho_t]_{11}$. \hfill $\blacksquare$

\section{Relation between bath correlation and spectral density functions}
\label{sectbathcorr}

In this section, we derive the relation between the spectral density and the bath correlation functions, see \fer{+4}. As the spectral density function is defined in the physics literature for environments of harmonic oscillators labelled by a discrete parameter (momentum), we first identify all quantities of our (continuous momentum, i.e., infinite volume) model with the equivalent discrete counterparts. 

In the seminal paper \cite{Leggett} on the spin-Boson system, the following model is considered. A spin is coupled to a bath of oscillators. The Hamiltonian of a single Hamiltonian, labelled by $\alpha$, is $H_\r$, as introduced before \fer{sbh}. The interaction of the bath oscillators with the spin is given by
\begin{equation}
\frac 12 q_0\sigma_z\sum_\alpha c_\alpha x_\alpha,
\label{+0}
\end{equation}
where $\sigma_z$ is the Pauli operator,
$$
x_\alpha =\frac{1}{\sqrt{2m_\alpha \omega_\alpha}}(a_\alpha+a^\dagger_\alpha),
$$ 
see equation (1.4) and p.7 of \cite{Leggett}. Here, $q_0$ is a coupling constant, $c_\alpha$ are real numbers and $a_\alpha, a^\dagger_\alpha$ are annihilation and creation operators, satisfying $a_\alpha a^\dagger_\beta-a^\dagger_\beta a_\alpha=\delta_{\alpha\beta}$ (Kronecker symbol). Leggett et al. define {\em the spectral density} in \cite{Leggett}, equation (1.5), by
\begin{equation}
J(\omega) = \frac\pi2\sum_\alpha \frac{c^2_\alpha}{m_\alpha\omega_\alpha} \delta(\omega-\omega_\alpha).
\label{specden}
\end{equation}
Our interaction in \fer{n0} has the form
\begin{equation}
\lambda \left[
\begin{array}{cc}
\gd & a\\
a & \ga
\end{array}
\right]\otimes\varphi(h).
\label{int}
\end{equation}
This interaction is identified with the quantities in Leggett et al.'s work by making a {\em discretization of momentum space} of the free field in a box of (very large) side length $L$:
\begin{eqnarray*}
\int_{{\mathbb R}^3}\d^3k &\sim& \left(\frac{2\pi}{L}\right)^3\sum_{k\in \frac{2\pi}{L}{\mathbb Z}^3}\\
a^\dagger(k) &\sim& \left(\frac{2\pi}{L}\right)^{\!\! -3/2} a^\dagger _k
\end{eqnarray*}
Then, taking the function $h(k)$ in \fer{int} to be real valued, we have
$$
\varphi(h) = \frac{1}{\sqrt{2}}\int_{{\mathbb R}^3}h(k)\{a^\dagger(k) +a(k)\}\d^3k \sim \frac{1}{\sqrt{2}}\left(\frac{2\pi}{L}\right)^{3/2}\!\!\!\sum_{k\in \frac{2\pi}{L}{\mathbb Z}^3} h_k \ (a^\dagger_k +a_k).
$$
Thus our interaction \fer{int} is of Leggett et al.'s form \fer{+0}, under the following identifications:
\begin{eqnarray}
a&=&0\label{id1}\\
\gd=-\ga&=&1/2\\
\lambda&=& q_0\\
\left(\frac{2\pi}{L}\right)^{3/2}\sum_{k\in \frac{2\pi}{L}{\mathbb Z}^3} &=&\sum_\alpha\\
\frac{h_k}{\sqrt{2}} &=&\frac{c_\alpha}{\sqrt{2m_\alpha\omega_\alpha}}.
\label{id2}
\end{eqnarray}

\bigskip

We evaluate the Fourier transform of the correlation function,
\begin{equation}
\widehat C(\omega) = \frac{1}{\sqrt{2\pi}}\int_{-\infty}^\infty\e^{-\i \omega t}\langle\varphi(t)\varphi\rangle \d t.
\label{ftcorr}
\end{equation}
We have
\begin{eqnarray}
\av{\varphi(t)\varphi}&=& \frac12\sum_{\alpha,\alpha'} h_\alpha h_{\alpha'} \av{(\e^{\i\omega_\alpha t}a^\dagger_\alpha+\e^{-\i t\omega_\alpha}a_\alpha)(a^\dagger_{\alpha'}+a_{\alpha'})}\nonumber\\
&=&\frac12 \sum_\alpha h^2_\alpha( \e^{\i\omega_\alpha t} n_\alpha +\e^{-\i\omega_\alpha t} (n_\alpha+1)),
\label{phitphi}
\end{eqnarray}
where $n_\alpha = \av{a^\dagger_\alpha a_\alpha}$ is the average occupation of mode $\alpha$. Taking the Fourier transform of \fer{phitphi}, and using that
\begin{equation}
\int_{-\infty}^\infty \e^{\i (\omega-\omega')t} \d t=2\pi \delta(\omega-\omega'),
\label{+9}
\end{equation}
we obtain, for $\omega\in\mathbb R$,
\begin{equation}
\widehat C(\omega) +\widehat C(-\omega) = \sqrt{\pi/2}\sum_\alpha h_\alpha^2 (1+2n_\alpha)\delta(\omega-\omega_\alpha).
\label{+3}
\end{equation}
{} For a thermal reservoir, we have $n_\alpha =(\e^{\beta \omega_\alpha}-1)^{-1}$, so
\begin{equation}
\widehat C(\omega) +\widehat C(-\omega) = \sqrt{2/\pi}\coth(\beta\omega/2)J(\omega)
,
\label{+4}
\end{equation}
where $J(\omega)$ is the spectral density \fer{specden}.

\section{Recovering Leggett et al.'s results for small tunneling}
\label{recleggsect}

In the setting of the Marcus theory of electron transport \cite{Marcus1,XuSch} the system-envoronment interaction is diagonal,  $a=0$ in \fer{n0}, and the tunneling element $V$ in \fer{n0} is considered to be very small. Transfer rates are then obtained perturbatively in orders of $V$. In this section, we show that the rate obtained in our paper is the same expression as that obtained in Leggett et al.'s work \cite{Leggett}.

{}For $a=0$ and $|V|<\!\!<\ed-\ea$ and to lowest order in $\lambda$ and $V$, our relaxation rate \fer{++5} is 
\begin{equation}
\gamma_{\rm relax} = 2\lambda^2 V^2 \left(\frac{\gd-\ga}{\ed-\ea}\right)^2\coth\big(\beta(\ed-\ea)/2\big) J(\ed-\ea).
\label{+++5}
\end{equation}
Since $\Omega=\ed-\ea+O(V)$, we have replaced $\Omega$ in \fer{++5} by the energy difference $\ed-\ea$ in \fer{+++5}. To compare our rate with Leggett's, we first harmonize notation. Our parameters are identified with those of the spin-boson model in \cite{Leggett} according to \fer{id1}-\fer{id2} and
\begin{eqnarray}
V&=&-\Delta/2\\
\ed-\ea&=&\epsilon,
\end{eqnarray}
where $\Delta$ and $\epsilon$ are the energy gap and the tunnelling constant of \cite{Leggett}. Therefore, in Leggett's notation, our transfer rate \fer{+++5} is given by 
\begin{equation}
\gamma_{\rm relax} = \frac{q_0^2}{2}\frac{\Delta^2}{\epsilon^2}\coth(\beta\epsilon/2) J(\epsilon).
\label{+10}
\end{equation}

\medskip
Leggett et al. use the heuristic ``golden rule'' approach (Section III.D of \cite{Leggett}), which is a formal perturbation theory in the tunneling constant $\Delta$ (plus some further approximations), to derive the following expression for the electron transfer rate (see equation (3.38) of \cite{Leggett})
\begin{equation}
\tau^{-1}=\Delta^2\int_0^\infty \d t\cos(\epsilon t)\cos[(q^2_0/\pi)Q_1(t)] \exp-[(q^2_0/\pi)Q_2(t)],
\label{tau}
\end{equation}
where
\begin{eqnarray}
Q_1(t)&=& \int_0^\infty\frac{J(\omega)}{\omega^2}\sin\omega t \ \d\omega,\\
Q_2(t)&=& \int_0^\infty\frac{J(\omega)(1-\cos\omega t)}{\omega^2}\coth(\beta\omega/2) \d\omega.
\label{+8}
\end{eqnarray}
Note that \fer{tau} contains all orders in $q_0$, but our result only yields the second order (i.e., the $q_0^2=\lambda^2$ term). While we use a mathematically rigorous perturbation theory, Leggett et al.\! proceed as follows. First, they apply a suitable unitary transformation to the spin-boson Hamiltonian, transforming it into (\!\cite{Leggett}, equation (3.30))
\begin{equation}
\widehat H' = -\frac12\Delta(\sigma_+\e^{-\i\Omega} +{\rm H.c.}) +\frac12\epsilon\sigma_z +\sum_\alpha\frac12(m_\alpha\omega_\alpha^2x_\alpha^2 +p_\alpha^2/m_\alpha).
\label{+7}
\end{equation}
Here, $\sigma_+=\frac12 (\sigma_x+\i\sigma_y)$ and $\Omega =\sum_\alpha(q_0c_\alpha/m_\alpha\omega^2_\alpha)p_\alpha$. For $\Delta=0$ this operator is explicitly diagonalized. Then they use perturbation theory in \fer{+7} for small $\Delta$, and obtain \fer{tau} to second order in $\Delta$. Note that in applying perturbation theory to \fer{+7}, the perturbation is $-\frac12\Delta(\sigma_+\e^{-\i\Omega} +{\rm H.c.})$, {\it which depends on $q_0$}, indirectly via $\Omega$. Even though this perturbation is bounded by ${\rm const.}|\Delta|$, independently of $q_0$, the approximations made to arrive at \fer{tau} are complicated, and it is not clear if \fer{tau} is the correct expression for relatively large $q_0$. (The correction term could be, for instance, of order $\Delta^4q_0^2$, which would then dominate the main part, \fer{tau}, for $\Delta^2 q_0^2>1$.) For small values of $q_0$, the perturbation series can be controlled, and formula \fer{tau} should be rigorously correct.

Expanding \fer{tau} for small $q_0$, we have
\begin{equation}
\tau^{-1} = \Delta^2 \int_0^\infty \d t\ \frac{\e^{\i\epsilon t}+\e^{-\i \epsilon t}}{2}\left[ 1-\frac{q^2_0}{\pi} Q_2(t)\right] +O(\Delta^2q_0^4).
\end{equation}
Using equation \fer{+8} we can integrate explicitly over the variable $t$ (c.f. \fer{+9}),
\begin{eqnarray}
\tau^{-1} &=& \frac{\Delta^2q_0^2}{2}\int_0^\infty \frac{J(\omega)}{\omega^2}\coth(\beta\omega/2)\delta(\omega-\epsilon)\d\omega +O(\Delta^2q_0^4)\nonumber\\
&=&  \frac{\Delta^2 q_0^2}{2\epsilon^2}J(\epsilon)\coth(\beta\epsilon/2)+O(\Delta^2q_0^4).
\label{+11}
\end{eqnarray}
Comparing \fer{+11} and \fer{+10} shows that:

\medskip
{\em 
The expression for the transfer rate obtained by the resonance method, to second order in the tunneling matrix element ($V$) and second order in the interaction between the spin and the reservoir ($\lambda$), is the same as that quantity obtained by the heuristic ``golden rule'' method used in Leggett et al. \cite{Leggett}. Note that the equality of the two expresssions holds for arbitrary spectral density functions $J$.}

\medskip

{\bf Remarks.\ }
 {\bf (1)} \cite{Leggett} gives an expression for the transfer rate which contains all orders in the coupling $q_0$ to the reservoir, see \fer{tau}. However, its derivation is not controlled: it is not known if the `remainders' in the perturbation arguments are smaller than the `main terms'. Our method is mathematically rigorous, giving bounds on all remainder terms in the perturbation theory arguments, in the parameter regime where the coupling to the reservoir is small enough. This means that our main term is {\it guaranteed} to be larger than the perturbation corrections. Higher order terms in the spin-reservoir coupling can be calculated rigorously with the resonance method, but their derivation is lengthy, and we did not check if they coincide with Leggett et al.'s expressions.

{\bf (2)} It has been shown in \cite{XuSch} that the spin-boson system (the same as in \cite{Leggett}) gives the same reaction rate for large temperatures (in the physiological regime). Since we have just shown that our results coincide with those of \cite{Leggett}, we have that {\em the resonance theory produces the same transfer rate as the Marcus theory, in the mathematically controllable parameter regime (small tunnelling matrix element) and at physiological temperatures}.

\bigskip
\bigskip

{\bf Acknowledgements.\ } M.M. acknowledges the support of the Natural Sciences and Engineering Research Council of Canada (NSERC) through an Individual Discovery Grant, and he is grateful for support from the Institut Henri Poincar\'e (IHP) through the programme ``Research in Paris''. The work by G.P.B. and R.S. was carried out under the auspices of the National Nuclear Security Administration of the U.S. Department of Energy at Los Alamos National Laboratory under Contract No. DE-AC52-06NA25396.

\end{document}